\documentclass[a4paper]{article}
\usepackage{amsmath,graphicx}
\usepackage{booktabs,multirow}

\usepackage{array}
\newcolumntype{P}[1]{>{\centering\arraybackslash}p{#1}}
\newcolumntype{M}[1]{>{\centering\arraybackslash}m{#1}}

\usepackage{INTERSPEECH2022}

\title{Towards Cross-speaker Reading Style Transfer on Audiobook Dataset}

\name{Xiang Li$^{1,\dagger}$, Changhe Song$^{1,\dagger}$, \thanks{$\dagger$ Equal contributions.} Xianhao Wei$^{1,3}$,  Zhiyong Wu$^{1,2,*}$, \thanks{* Corresponding author.} Jia Jia$^{1,3}$, Helen Meng$^{1,2}$}

\address{
  $^1$Tsinghua-CUHK Joint Research Center for Media Sciences, Technologies and Systems,\\Shenzhen International Graduate School, Tsinghua University, Shenzhen, China\\
   $^2$Department of Systems Engineering and Engineering Management,\\The Chinese University of Hong Kong, Hong Kong SAR, China \\
   $^3$Department of Computer Science and Technology, Tsinghua University, Beijing, China
}

\email{
\{xiang-li20, sch19, weixh20\}@mails.tsinghua.edu.cn, \{zywu, hmmeng\}@se.cuhk.edu.hk, jjia@tsinghua.edu.cn}

\begin{document}

\maketitle
\begin{abstract}
Cross-speaker style transfer aims to
extract the speech style of the given reference speech,
which can be reproduced
in the timbre of arbitrary target speakers.
Existing methods on this topic have explored
utilizing utterance-level style labels
to perform style transfer
via either global or local scale style representations.
However, audiobook datasets are typically
characterized by both the local prosody and global genre,
and are rarely accompanied by utterance-level style labels.
Thus, properly transferring the reading style across different speakers remains a challenging task.
This paper aims to introduce a chunk-wise multi-scale cross-speaker style model
to capture both the global genre and the local prosody in audiobook speeches.
Moreover, by disentangling speaker timbre and style
with the proposed switchable adversarial classifiers,
the extracted reading style is made adaptable
to the timbre of different speakers.
Experiment results confirm that the model manages to transfer a given reading style to new target speakers.
With the support of local prosody and global genre type predictor,
the potentiality of the proposed method in 
multi-speaker audiobook generation is further revealed.

\end{abstract}
\section{Introduction}
\label{sec:intro}


\begin{figure*}[t]
  \centering
  \includegraphics[width=0.9\linewidth]{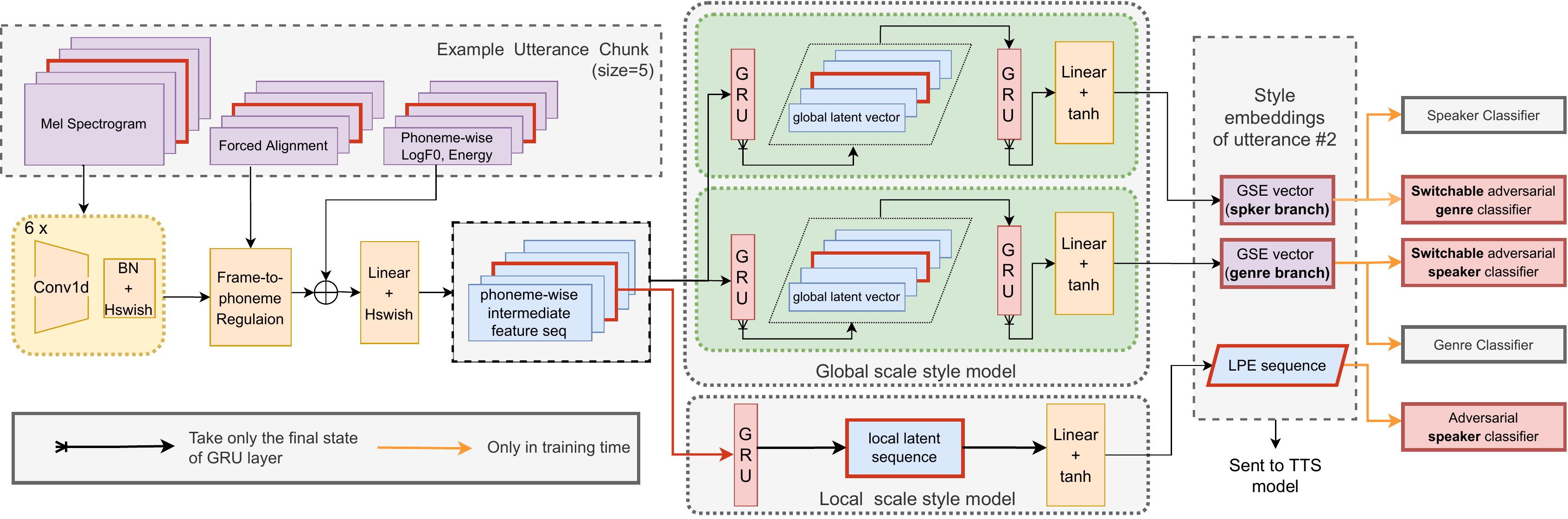}
  \caption{Chunk-wise multi-speaker multi-scale reference encoder with 2 global branches}
  \label{fig:chunk}
\end{figure*}

The past decade has witnessed a flourishing development in the area of neural text-to-speech (TTS)
\cite{Wang2017,shen2018natural,DBLP:conf/nips/KongKB20}.
Recently, expressive TTS is receiving growing attention \cite{skerry2018towards,wang2018style,lee2019robust,Klimkov2019},
since style variations are important for synthesizing more natural speeches.
However, in the real world, high-quality expressive datasets
typically contain only a small number of speakers due to expense concerns.
To generate stylized speeches of arbitrary speakers,
it is necessary to transfer the given speech style to the target speakers,
whose recorded audio data exclude the target style.

Existing studies on cross-speaker style transfer basically follow
the practice of extracting style information
from given reference speeches,
which is widely adopted in expressive TTS \cite{skerry2018towards,wang2018style}.
The extracted style is subsequently organized into
either a global style embedding \cite{bian2019multi,DBLP:conf/interspeech/WhitehillMMS20},
or a fine-grained local style embedding sequence \cite{xie2021multi},
which are supposed to be adaptable to various speakers. 
Specifically, to distinguish the characteristics of the target style,
utterance-level style labels including prosody class and emotion category
are usually utilized.

A part of existing works on this topic
focus on transferring various prosody classes
(e.g. poetry reading, call-center)
across different speakers:
the multi-reference Tacotron \cite{bian2019multi}
implements a method based on multiple Global Style Tokens (GST),
where speaker timbre and prosody class are embedded into different
global style dimensions and 
disentangled via inter-cross training strategy;
\cite{xie2021multi} attaches a fine-grained prosody module to the
multi-speaker Tacotron-2 \cite{shen2018natural},
which extracts local prosody embedding as an additional input of the decoder.
Hence, prosody transfer is directly realized
by switching speaker embedding
while keeping the local prosody representation.

Other studies intend to transfer given speech emotions (e.g. happy, angry) to speakers with only neutral data:
\cite{DBLP:conf/interspeech/WhitehillMMS20} employs multiple reference audios to provide speaker and emotion embeddings separately, which is achieved by training on 
$<${text, style-matched audio, GT audio}$>$ and $<$text, random audio, random audio$>$ triplets with adversarial cycle consistency;
\cite{li2021controllable} introduces an emotion disentangling module (EDM) for the disentanglement of speaker timbre and emotion style attributes.
Recently proposed methods further insert conditional layer normalization into the TTS decoder for better adaptation performance,
to which the target speaker ID and desired emotion labels are fed as input condition \cite{liu2021meta,wu2021cross}.

Compared with prosody-class-labeled or emotional corpus,
the reading style of audiobook datasets is usually observed to
emerge on both local and global scales:
On the one hand, the rich local prosody variation in speech
is one of the crucial elements to make the content of an audiobook
more attractive to listeners and easier to follow
(e.g. there are often prosody changes at the 
joint point between character lines and narrator lines).
On the other hand, the expressiveness
of audiobook datasets is also
characterized by the steady global genre,
which must fit the topic of the original document
(e.g.
fairy tales are supposed to be read in an innocent tone with patience).
Since the type of global genre is determined by the book topic,
the same genre label is shared across the whole book,
which could be directly obtained according to the book content,
rather than being manually annotated sentence by sentence like
emotion.

However, the fixed document-level genre label rarely contains
the particular style information of a specific utterance,
which makes it difficult to directly apply existing
utterance-label-based style model to audiobook datasets.
The multi-scale characteristics of audiobook reading style also
bring challenges to utilize the aforementioned single-scale
cross-speaker style transfer methods. 
With these concerns,
this paper adopts the multi-scale style modeling scheme in \cite{li21r_interspeech},
and proposes a chunk-wise multi-speaker multi-scale reference encoder
to model speaker timbre and global genre
of the audiobook on the global scale,
while fine-grained prosody is modeled on the local scale.
Specifically, the speaker timbre and genre are extracted
from a chunk of consecutive utterances,
since a larger view would help the model capture these factors.
In order to disentangle reading style and timbre,
a novel adversarial training strategy based on
switchable adversarial classifiers are devised to
provide style transfer ability among disjoint datasets,
which is the common training basis of cross-speaker reading style transfer in reality.


\section{Methodology}
\label{sec:meth}
We construct an expressive TTS system with Tacotron-2 \cite{shen2018natural}
as backbone.
The integrated multi-scale cross-speaker style model consists of two components:
(i) A chunk-wise multi-speaker multi-scale reference encoder;
(ii) Switchable adversarial classifiers for speaker timbre and style disentanglement.

\subsection{Multi-scale cross-speaker expressive TTS scheme}
\label{ssec:meth:ms}
The proposed model adopts the idea of extracting a global-scale style embedding (GSE) vector
and a local-scale prosody embedding (LPE) sequence
from the given reference speech
with a multi-scale reference encoder \cite{li21r_interspeech}, 
which represent the global genre and local prosody of audiobook reading style, respectively.
The LPE sequence and the repeated GSE vector
are then attached to the text embedding to stylize the generated speech.

For cross-speaker style modeling,
a multi-branch global style module is introduced
to our reference encoder,
which is able to extract two disentangled GSE vectors based on two branches.
One of the GSE vectors is assigned to model speaker timbre
since timbre is commonly recognized as a coarse style attribute.
And the other is set to model
the global genre of the audiobook.
As a result,
cross-speaker style transfer could be achieved by switching speaker timbre GSE vector.

\subsection{Chunk-wise multi-speaker multi-scale reference encoder}
\label{ssec:meth:chunk}

The workflow of our reference encoder is divided into two steps.
The input mel-spectrograms of reference speeches are first fed into 6 1D-convolution layers,
which share the same structure as those in \cite{li21r_interspeech}
except that Hardswish \cite{howard2019searching} is adopted as the activation function
instead of ReLU for better performance.
The produced frame-wise feature sequence of each utterance
is then regularized into phoneme-wise
by averaging frames within the same phoneme,
according to the prepared forced alignment results.
Eventually, the regularized sequence is concatenated with
preprocessed phoneme-wise acoustic features (logF0 and energy)
and fed through a linear layer with Hardswish activation
to obtain the final phoneme-wise intermediate feature sequence.

During the second step, the intermediate sequence is shared for both global and local scale modules as input:

On the local scale, 
the sequence is sent through a GRU layer, followed by a linear layer and tanh activation,
to get the pre-aligned LPE sequence.
This output sequence is limited to a small number of channels (4 in our setting)
by the linear layer to form an information bottleneck,
which is typically adopted to tackle content or timbre leakage problems.

On the global scale, style embedding is extracted on a chunk-wise basis.
The definition of a chunk is 
a short paragraph in the audiobook made up of
a couple of consecutive utterances.
Compared with single-utterance style models,
the chunk-wise model has access to a larger scope
and generates 
a smoother representation of the global genre of the audiobook,
since style fluctuation among different utterances is averaged. 
Specifically, every branch shares the same
network structure and workflow:
Each intermediate feature sequence 
is first compressed into a global vector
by passing through a GRU layer and taking the final state as output.
All global vectors 
within the same chunk are then consecutively stacked together as a sequence,
and compressed into a vector in the same way by another GRU.
The final state of the GRU is 
processed by a linear layer and tanh activation,
whose
output is the eventual GSE vector on the current branch,
and is shared across the whole chunk.

\begin{figure}[t]
  \centering
  \includegraphics[width=0.7\linewidth]{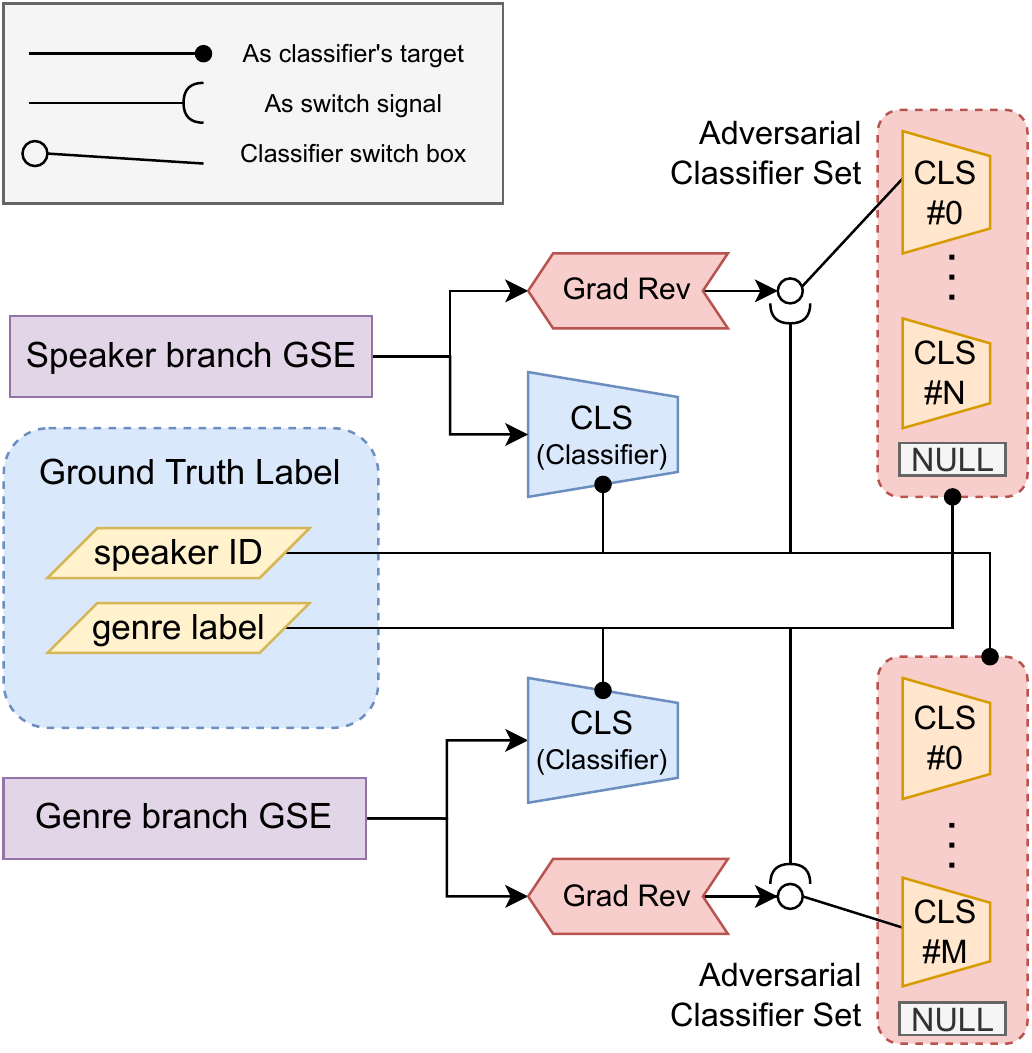}
  \caption{Adversarial training with switchable classifiers}
  \label{fig:grl}
  \vspace{-0.5cm}
\end{figure}

\subsection{Cross-speaker training strategy}
\label{ssec:meth:train}
\subsubsection{Two-stage training process}
The multi-scale cross-speaker style model is trained in a two-stage fashion
to eliminate interference 
among different scales.

During the first stage, the global-scale module is excluded.
The model focuses on learning a speaker-agnostic local scale prosody representation
by 
sending the LPE sequences to an adversarial speaker classifier
which consists of a gradient reverse layer and 2 linear layers.
Meanwhile, similar to the setting in \cite{xie2021multi},
a temporary speaker embedding table is used
to generate speaker embeddings
that are attached to the text embedding
to provide speaker timbre information for the TTS decoder.

During the second stage, the temporary speaker embedding table is dropped;
the 1D-convolution layers, local scale style module,
and the text encoder of Tacotron-2 are frozen;
the decoder of Tacotron-2 is reinitialized.
The model is committed to extracting disentangled timbre/genre features 
by feeding each GSE vector through corresponding
speaker/genre label classifiers
and switchable adversarial genre/speaker classifiers.

\subsubsection{Switchable adversarial classifiers}
As shown in Figure.\ref{fig:grl},
different from vanilla gradient-reverse-layer-based 
adversarial classifier,
the proposed switchable adversarial classifier (SAC)
introduces: (i) multiple underlying classifiers,
each of which is made up of 2 linear layers;
(ii) an additional switch box to choose the proper underlying classifier for each input sample
according to its corresponding label
(speaker ID on speaker branch, genre label on genre branch).
Consequently, there is a unique underlying classifier for each corresponding label of the branch, except for those with only one prediction target (e.g. on the speaker branch, those would be the speakers with only one type of genre), 
none of the classifiers is applied (noted as \textit{NULL} in Figure.\ref{fig:grl}).

The switchable adversarial classifiers are designed to
accommodate the common scenario of
cross-speaker reading style transfer task:
training on disjoint datasets.
More specifically,
cross-speaker reading style transfer is usually conducted among
several audiobooks with different narrators and topics.
Which indicates that
each speaker may have different types of global genre, if more than one;
whereas each genre label may correspond to various numbers of speakers.
Thus, it is more proper to
assign a specific adversarial classifier for each label, rather than depending on the same classifier to handle
the various data circumstances of all labels.
\section{Experiments}
\label{sec:expr}
\subsection{Datasets and model details}
\label{ssec:expr:dataset_model}
We evaluate the proposed cross-speaker reading style transfer method on the following 
disjoint datasets:

\textbf{MST-Originbeat}: A neutral Mandarin corpus offered by the ICASSP 2021 M2VoC challenge \cite{9414001},
with one female and one male speaker,
each with 5,000 utterances.

\textbf{DB}: A private neutral Mandarin dataset
with 10,000 utterances from
another female Chinese speaker, named DB6.

\textbf{Audiobook\_FM}: A private Mandarin audiobook dataset with 8 speakers and 2 types of genres:
fairy tale and Chinese martial arts fiction.
A female and a male speakers cover both of the 2 genres,
each of the other 6 speakers only covers one of the 2 genres.
One of the 6 speakers is DB6, who only reads the fairy tale documents.
This dataset has a total amount of 1,315 paragraphs,
which is 13,718 utterances adding up to 24.3 hours.

When applied to the proposed chunk-wise reading style model,
each short paragraph in the Audiobook\_FM dataset is regarded as a chunk,
whose global genre label is assigned based on the topic of the including document:
either \textbf{fairy tale} or \textbf{martial arts fiction}.
While for datasets MST-Originbeat and DB,
each chunk is made up of 10 randomly sampled utterances
that are voiced by the same speaker,
and all chunks share the same global genre label: \textbf{neutral}.


In our experiment,
the model is trained 380k steps for the first stage,
260k steps for the second stage.
All classifiers are trained by cross-entropy loss with weight set as $0.05$.
A pre-trained HiFi-GAN \cite{DBLP:conf/nips/KongKB20} vocoder
is utilized to generate audio.

\subsection{Cross-speaker reading style transfer}
\label{ssec:expr:style_trans}

Based on the proposed model,
we aim to transfer the reading style of
fairy tale and martial arts fiction to neutral speakers in MST-Originbeat,
or speakers in whose training data the target reading style is absent.
Given the target speaker identity and the audios of the reference paragraph with the target reading style,
this could be achieved by combining:
(i) the averaged speaker timbre GSE vector
over all chunks in the training data of the target speaker;
(ii) the global genre GSE vector and LPE sequences
extracted from the reference audio chunk.
We evaluate the reading style transfer results on the reserved test set,
the proposed synthesized speech keeps local prosody and global genre
close to the reference utterance,
but with the timbre of target speakers.
For audio samples,
please check our demo website\footnote{\scriptsize https://thuhcsi.github.io/is2022-cross-speaker-reading-style-transfer}.

\subsubsection{Baseline}
For comparison, we establish an embedding-table-based baseline method by
replacing the 2 global branches of the multi-scale style model with
a speaker embedding table and a global genre embedding table.
Which is similar to the setting of \cite{xie2021multi},
except that there is an additional embedding table of the global genre
to accommodate the audiobook dataset.

\subsubsection{Evaluation}
We conduct a Mean Opinion Score (MOS) test on 
66 utterances from the test set,
with 26 native Chinese speakers serving as subjects.
Each test group is made up of a synthesized audio $\tilde{M}$,
the ground truth reference audio $M_{ref}$ and the target speaker audio $M_{tgt}$,
The subjects are asked to rate $\tilde{M}$
on a scale of 1 to 5 regarding 3 different aspects:
(i) its style similarity to $M_{ref}$;
(ii) its timbre similarity to $M_{tgt}$;
(iii) its audio quality.
As shown in Table.\ref{tab:mos},
the proposed model beats the baseline model on
all evaluation scores,
which verifies the effectiveness
of the proposed chunk-wise multi-scale cross-speaker style model
in enhancing cross-speaker reading style transfer performance.

In addition to the subjective timbre similarity test,
we use a pre-trained speaker verification (SV) model \cite{zhang2022mfa} to 
extract speaker embedding vectors from the synthesized audios and the target speaker audios.
The objective evaluation of speaker timbre similarity of each synthesized speech is thus obtained
by computing the cosine similarity between the extracted vector
and the averaged vectors of the corresponding target speaker.
As shown in Table.\ref{tab:mos},
the overall outcomes are generally consistent with subjective timbre similarity test results.

\begin{table*}[th]
    \centering
    \caption{Similarity/Quality MOS and SV embedding similarity results (Average score and 95\% confidence interval)}
    \label{tab:mos}
    \begin{tabular}{m{7.8em}|ccc|c|c|c}
        \toprule
         \multirow{2}{7.8em}{Models} & \multicolumn{3}{c|}{Style Similarity} & Timbre & SV embedding & Audio \\
         & Fairy tale & Maritial arts fiction & Overall & Similarity & Cosine similarity & Quality Score \\
        \midrule
        Baseline & $3.57\pm0.10$ & $3.99\pm0.10$ & $3.79\pm0.07$ & $2.89\pm0.08$ & $0.75\pm0.02$ & $3.08\pm0.10$ \\
        Proposed & $\mathbf{3.88\pm0.08}$ & $\mathbf{4.02\pm0.0}$ & $\mathbf{3.95\pm0.06}$ & $3.22\pm0.07$ & $0.76\pm0.02$ & $3.58\pm0.06$ \\
        \hspace{3mm}w/o GSE.style & $3.79\pm0.09$ & $3.80\pm0.10$ & $3.79\pm0.07$ & $\mathbf{3.49\pm0.07}$ & $\mathbf{0.81\pm0.02}$ & $3.56\pm0.06$ \\
        \hspace{3mm}w/o chunk & $3.81\pm0.08$ & $3.80\pm0.10$ & $3.80\pm0.06$ & $3.44\pm0.07$ & $0.79\pm0.02$ & $\mathbf{3.60\pm0.06}$ \\
        \hspace{3mm}w/o SAC & $3.67\pm0.10$ & $4.01\pm0.11$ & $3.85\pm0.07$ & $2.83\pm0.09$ & $0.48\pm0.06$ & $3.49\pm0.07$ \\
        \midrule
        GT & - & - & - & - & - & $4.32\pm0.09$ \\
        \bottomrule
    \end{tabular}
  \vspace{-0.3cm}
\end{table*}

\begin{figure}[th]
  \centering
  \includegraphics[width=0.9\linewidth]{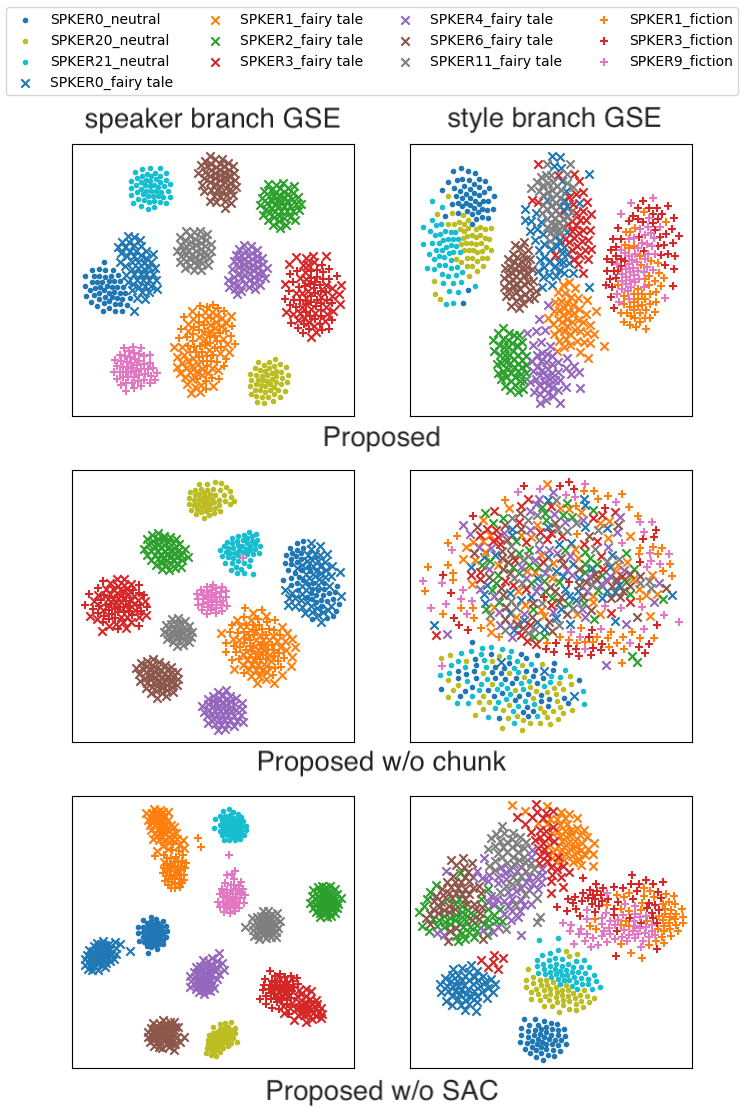}
  \caption{t-SNE plot for speaker GSE vectors \textit{(left column)} and genre GSE vectors \textit{(right column)}. Marker colors correspond to speaker IDs; marker shapes correspond to genre labels. The proposed model reaches the best clustering on both branches.}
  \label{fig:t-sne}
  \vspace{-0.5cm}
\end{figure}


\subsection{Ablation study}
\label{ssec:expr:abl}
We conduct 3 settings of ablation experiments to reveal the functionality of each component of the proposed method.

(i) We remove the global genre branch in the style model,
leaving the timbre branch as the only global module.
As shown in the subjective test results,
there is noticeable style similarity degradation
in the transferred speech of the ablating model,
which indicates the necessity of explicitly modeling
the global genre in reading style transfer.
In the meantime, 
the timbre similarity scores are observed to be improved,
since on the global scale, the ablating model is focused on speaker timbre only.

(ii) We replace the chunk-wise GSE extracting method
with an ordinary utterance-wise approach.
The style similarity evaluation results are downgraded
due to the shrunken receptive field of the global scale modules.
After visualizing the extracted GSE vectors via t-SNE,
the genre GSE vectors of the fairy tales are found to be confused
with those of the martial arts fiction
(right column of the 2nd row in Figure.\ref{fig:t-sne}).
On the other hand,
considering there are unique GSE vectors for each utterance,
more information is conveyed
to help the model converge better,
which explains the slight promotion in audio quality.

(iii) We employ vanilla adversarial classifiers
to disentangle GSE vectors of different branches,
instead of the proposed switchable adversarial classifiers (SAC).
As revealed in Table.\ref{tab:mos},
the speaker timbre similarity of transferred speeches
drops significantly.
Furthermore, in the t-SNE plot of the extracted timbre GSE vectors,
the samples of $<\mathrm{SPKER1\_fiction}>$ and $<\mathrm{SPKER1\_fairy\ tale}>$
end up in different clusters
(left column of the 3rd row in Fig.\ref{fig:t-sne}),
which confirms there is still considerable genre information
entangled in the timbre GSE vector.
These details support that the proposed SAC is crucial for
timbre and style disentanglement on disjoint dataset scenarios.

\subsection{Automatic audiobook generation}
Based on the proposed cross-speaker reading style transfer model,
an automatic audiobook generation system could be constructed
by incorporating a text analysis model
which predicts the LPE and genre from given book content.
In our practice,
the prediction model is implemented with RNN and linear layers,
which takes BERT \cite{devlin2018bert} token embedding and Tacotron-2 phoneme embedding as its inputs,
similar to existing methods \cite{hodari2021camp,xie2021multi}.
According to the predicted genre label and the identity of the desired speaker,
the GSE vectors on each branch could be obtained
by choosing the averaged GSE vectors over the training data of the target genre/speaker.
Together with the predicted LPE and text sequence,
the speeches of the target speaker reading the material with the predicted style
is eventually generated.
The inference results are presented on our demo site.

\label{ssec:expr:abl}
\section{Conclusions}
\label{sec:conclusions}

In this work,
a chunk-wise multi-scale cross-speaker speech style model is introduced
for cross-speaker reading style transfer task on audiobook dataset.
Experiments on disjoint corpus indicate improvements
on both speaker and style similarities compared to the baseline model.
Ablation study reveals the necessity of explicitly modeling
the global genre of audiobook on a chunk basis;
and the switchable adversarial classifiers
are verified to be effective in style embedding disentanglement.
In the future, we intend to generalize the proposed method
to larger-scale multi-speaker TTS corpus,
and improve the speaker disentanglement on local scale prosody representation.

\textbf{Acknowledgement} This work is supported by National Key R\&D Program of China (2020AAA0104500), National Natural Science Foundation of China (62076144) and Shenzhen Key Laboratory of next generation interactive media innovative technology (ZDSYS20210623092001004).






\bibliographystyle{IEEEtran}
\bibliography{template}


\end{document}